\documentclass[3p,twocolumn]{elsarticle}
\usepackage{amsmath,amssymb,amsfonts,bbm,graphicx,times,color,mathptmx,hyperref}
\usepackage{mathrsfs}

\newcommand{\bmsigma}{\boldsymbol \sigma}

\newcommand{\bmA}{\boldsymbol A} 
\newcommand{\bmC}{\boldsymbol C}

\begin{document}
\title{Continuous-variable entanglement dynamics in Lorentzian environment}
\author{Berihu Teklu}
\address{Department of Applied Mathematics and Sciences, Center for Cyber-Physical Systems (C2PS), Khalifa University, Abu Dhabi 127788, United Arab Emirates}

%\email{berihu.gebrehiwot@ku.ac.ae}
\date{\today}
%%%%%%%%%%%%%%%%%%%%%%%%%%%%%%%%%%%%%%%%%%%
\begin{abstract}
We address the non-Markovian entanglement dynamics for bimodal continuous variable quantum systems interacting with two independent structured reservoirs. We derive an analytical expression for the entanglement of formation without performing the 
Markov and the secular approximations.  We observe a variety of qualitative features such as entanglement sudden death, dynamical generation, and protection for two types of Lorentzian spectral densities, assuming the two modes initially excited in a twin-beam state. Our quantitative analysis shows that these cases with different reservoir spectrum, the environmental temperature and the initial amount of entanglement differ significantly in these qualitative features. 

\end{abstract}
%\pacs{03.67.Mn,03.65.Yz}

\maketitle

%%%%%%%%%%%%%%%%%%%%%%%%%%%%%%%%%%%%%%
\section{Introduction}\label{s:intro}
Recent advances in the study of the evolution of entanglement for open quantum systems
is an important issue for both fundamental and practical reasons, as quantum
entanglement is an essential resource for quantum information processing \cite{
NieChu,Horo09}.
The properties of open systems play an essential role in determining how the observed
classical world emerges from its quantum mechanical underpinnings via the process
known as decoherence. Any realistic analysis of quantum information protocols should
take into account the decoherence effect of the environment. In order to realize the
promise of a quantum information processor, the inevitable decoherence-inducing effect
of any system-environment interaction must be taken into account. The time scales on
which these processes take place and strategies to reduce these effects are a major topic
of research. Therefore, in order to describe the dynamics of the system of interest, the
theoretical results strongly depend on the underlying dissipative dynamics and the
performed approximations. Within the theory of open quantum systems, the dissipative
dynamics are mainly described by master equations of the reduced density matrix. The
most relevant Born-Markov approximation assumes weak coupling between the system
and the environment to justify a perturbative treatment and neglects short-time
correlations between the system and the reservoir. This approach has been widely and
successfully employed in the field of quantum optics \cite {Walls} where the
characteristic time scale of the environmental correlations is much shorter compared to
the internal system dynamics. There are a few cases, however, where an exact analytical
description of the dynamics is possible. Two relevant examples are the quantum
Brownian motion (QBM)~\cite{ Breuer,Weiss,Lind,Zurek,Lu} and the case of a two-level 
atom interacting with a zero-temperature reservoir with Lorentzian spectral density
\cite{ Breuer,Imamoglu,Garrya,Garry,Mazzola09,Mazzola,Richard}. 

\par
Any physical operations that reflects the time evolution of the state of a quantum system 
can be regarded as a channel. In particular, quantum channels grasp the way how 
quantum states are modified when subjected to noisy quantum communication lines. 
Couplings to other external degree of freedom, often beyond detailed control, will 
typically lead to losses and decoherence, effects that are modeled by appropriate 
non-unitary quantum channels.
\par
Quantum entanglement in continuous variables (CV) quantum channels has been
considered as a key resource in many aspects and applications of quantum information
processing due to potential improvement in the channel capacity \cite{Holevo}. In the
real world, the quantum coherence and entanglement of a quantum system will inevitably
be influenced and degraded by the external environment. There are several investigations
in the analysis of CV channels in the description of noisy CV quantum channels taking
into account decoherence and dissipation phenomena \cite{Serafini}. In those
investigations, the Markovian approximations or/and the rotating wave approximations 
(RWA) is/are assumed. However, if a short time interval or regime, comparable with the 
environment correlation time, is concerned, or if the environment is structured with a 
particular spectral density, then non-Markovian environment effect could become 
significant. For example, for high-speed quantum communication where the 
characteristic time scales become comparable with the reservoir correlation time. 
\par
Challenged by new experimental evidence a growing interest in non-Markovian 
descriptions has developed. From some phenomenological 
\cite{Ban,McAneney} and microscopic models \cite{Maniscalco,Jhn,Kl,Jhn1,Jhn2} of 
non-Markovian quantum channels have been proposed. Entanglement dynamics for such 
kinds of systems was analyzed before and a variety of results available. It was shown that 
the environment may completely disentangle an initial entangled state \cite{Maniscalco,Matteo, Serafini1}.
\par
A few theoretical proposals advance the entanglement dynamics in noisy 
continuous variable (CV) \cite{Braunstein}: most generally the interaction between the 
system-reservoir and a structured environment, where the system-reservoir correlations 
persist long enough to require a non-Markovian treatment, even in the weak coupling 
limit \cite{Maniscalco,Vasile2009, Vasile2010,Vasile2011,Bina2018}. Besides its own 
importance from a theoretical point of view, the role of structured environments and 
non-Markovianity in quantum metrology \cite{Alex,Claudia,Albarelli,Fahimeh,Dario,Gebbia,Candeloro,Ze}, 
quantum key distribution \cite{Vasile2011a}, and channel capacities \cite{Bylicka}, showing that non-Markovian 
quantum channels may be advantageous compared to Markovian ones. The theory 
developed in this article is relevant to the evolution of entanglement in bimodal continuous variable quantum 
systems interacting with two independent structured reservoirs. In this paper, we will analyze the 
entanglement dynamics using an analytic expression for the evolution of the 
entanglement of formation (EOF) \cite{EoF,G:03,EoF2008}. We assume the 
two oscillators initially excited in a twin-beam state (TWB). Despite its great simplicity, 
these models allow to describe the transition between Markovian and non-Markovian dynamics 
in experimental realization \cite{Liu,Tang}. By considering a Lorentzian environment 
we are able to obtained the exact reduced dynamics by means of the pseudomode 
method. 

\par
The outline of this article is as follows: in Sec. \ref{s:EME}, we describe our model and
present the non-Markovian master equation. Section \ref{s:EDynGS} presents an
overview about two-mode Gaussian states and the solution of the master equation with an
initial Gaussian state. In addition, we present the concept of EOF for two-mode CV
Gaussian state. In Sec. \ref{sec:ED} we describe cases where entanglement meets
entanglement sudden death (ESD)  and identify conditions or parameters where
entanglement can survive, some for ever - ``always alive'' (AL). Finally, in Sec.  \ref{s:concl} we 
conclude with a summary and outlook.

%%%%%%%%%%%%%%%
\section{Effective Master Equation and Weak Coupling}\label{s:EME}

The focus of our analysis is the non-Markovian dynamics of a system of two identical non-interacting 
quantum harmonic oscillators, each of them coupled to its own bosonic structured reservoir, giving 
rise to the total Hamiltonian    

\begin{eqnarray}
H&=&\sum_{j=1,2}\hbar\omega_0 a^{\dag}_j
a_j+\sum_{j=1,2}\sum_k\hbar\omega_{jk}
b^{\dag}_{jk} b_{jk} \nonumber \\
&+&\sum_{j=1,2}\sum_k
\gamma_{jk}(a_j+a_j^{\dag})(b_{jk}+b_{jk}^{\dag}).
\end{eqnarray}
In this Hamiltonian the first, second and third terms describe the system, environment and interaction, 
respectively, with $a_{j}$($a_{j}^{\dag}$) and $b_{jk}$ ($b_{jk}^{\dag}$) being the annihilation (creation) 
operators of the system and reservoirs harmonic oscillators, respectively. $\omega_0$ the oscillators 
frequency,  $\omega_{1k}$ and $\omega_{2k}$ the frequencies of the reservoirs modes, and 
$\gamma_{jk}$ the coupling between the $j$-th oscillator and the $k$-th mode of its environment. 
In the following we assume that the reservoirs have the same spectrum and are equally coupled 
to the oscillators.

The problem of a quantum harmonic oscillator interacting with a bosonic reservoir 
in thermal equilibrium has been derived for the first time in Ref. \cite{Lu}, and it is 
usually referred to as Hu-Paz-Zhang Master equation. The exact non-Markovian 
master equation for the bimodal field interacts bilinearly with  two identical uncorrelated 
bosonic thermal reservoirs takes the form \cite{Maniscalco}

\begin{align}
 \dot\varrho(t) =&  \sum_k \Big\{ \frac{1}{i \hbar} [H_k^0,\varrho (t)]
 -   \Delta(t) [X_k,[X_k,\varrho(t)]]  \nonumber \\
 &+ \Pi(t) [X_k,[P_k,\varrho(t)]]+ \frac{i}{2} r(t)
 [X_k^2,\varrho (t)] \nonumber \\
 &-  i \gamma(t) [X_k,\{P_k,\varrho\}]\Big\}, \label{QBMme}
\end{align}
where  
$$X_k = \frac{1}{\sqrt{2}} \left( a_k + a_k^{\dag}\right) \quad 
P_k=\frac{i}{\sqrt{2}} \left( a_k^{\dag}- a_k\right)$$ 
are the dimensionless quadrature operators. The operators $\hat a^{\dag}$ and 
$\hat a$ are the creation and annihilation operators of the two oscillators  and
$H_k^0= \hbar \omega_0 (a^{\dag}_k a_k+1/2)$ . These 
operators satisfy the bosonic commutation relation  $[a_k, a^{\dag}_k]=1$
($k=1,2$). We see that the Master equation given by  Eq.~(\ref{QBMme}),  
while it is exact, it describes also the non-Markovian system-reservoir 
correlations due to the finite correlation time of the reservoir.  All 
the non-Markovian character of the system is contained in the time 
dependent coefficients, $\Delta (t)$, $\Pi(t)$, $r(t)$ and $\gamma(t)$,
appearing in the Master equation. These coefficients depend
only on the reservoir spectral density, that is, on the microscopic 
effective coupling strength between the system
oscillator and the oscillators of the reservoir. The coefficient
$r(t)$ describes a time dependent frequency shift, $\gamma(t)$ is
the damping coefficient, $\Delta(t)$ and $\Pi(t)$ are the normal
and the anomalous diffusion coefficients, respectively
\cite{Breuer,Zurek}. It is worth noting that the Master equation given by 
Eq.~(\ref{QBMme}) is valid for general forms of the reservoir spectral density 
$J(\omega)$ and any temperature $T$. These time dependent coefficients are 
expressed in terms of series expansions in the system-reservoir coupling constant $\alpha$. 
Demanding weak system-reservoir couplings, that is, fulfilling the condition $\alpha\ll 1$,  
one can stop the expansion to the second order in the coupling constant and obtain 
analytic solutions for the coefficients. The analytic expression are given in the Appendix 
for the both low and high temperature-reservoirs characterized by a spectral densities of the 
form (\ref{den1}) and (\ref{den2}) respectively.

\par
The method is based on the quantum characteristic approach \cite{Intra2003} in 
which the solution of the master equation \eqref{QBMme} may be written as 
\begin{equation}\begin{split}\label{CharSol}
\chi_t(\Lambda)=\mbox{} &e^{-\Lambda^T
[\bar{W}(t)\oplus\bar{W}(t)]\Lambda}\\
&\times\chi_0(e^{-\Gamma(t)/2}[R^{-1}(t)\oplus R^{-1}(t)]\Lambda),
\end{split}\end{equation}
where $\chi_t(\Lambda)$
is the characteristic function at time $t$, $\chi_0$ is the
characteristic function at the initial time $t=0$,
$\Lambda = (x_1,p_1,x_2,p_2)$ is the two-dimensional phase space
variables vector, $\Gamma(t)=2\int_0^t\gamma(t')dt'$, and
$\bar{W}(t)$ and $R^{-1}(t)$ are $2 \times 2$ matrices.
The former matrix is given by
\begin{equation}\label{mat1}
\bar{W}(t)=e^{-\Gamma(t)}[R^{-1}(t)]^T W(t)R^{-1}(t),
\end{equation}
while the latter one, $R(t)$, contains rapidly oscillating terms.
In the weak coupling limit $R(t)$ takes the form
\begin{equation}\label{mat2}
R(t)=\left(
      \begin{array}{cc}
        \cos\omega_0 t & \sin\omega_0 t \\
        -\sin\omega_0 t & \cos\omega_0 t \\
      \end{array}
    \right).
\end{equation}
Finally, $W(t)=\int_{0}^{t}e^{\Gamma(s)}\bar{M}(s)ds$ with
$\bar{M}(s)=R^{T}(s)M(s)R(s)$ and
\begin{equation}\label{eq:Mt} 
M(s)=\left(
       \begin{array}{cc}
         \Delta(s) & -\Pi(s)/2 \\
         -\Pi(s)/2 & 0 \\
       \end{array}
     \right).
\end{equation}

%%%%%%%%%%
\section{Entanglement dynamics for Gaussian states}\label{s:EDynGS}
After the description of the master equation and its general solution 
through the characteristic function approach, we move on to a detailed 
analytic solution for the characteristic function in the weak coupling limit 
obtained in \cite{Intra2003}. Note first that the initial state of the subsystem 
is taken of Gaussian form and the evolution induced by the master equation 
\eqref{QBMme}  under the quantum dynamical semigroup assures the 
preservation in time of the Gaussian form of the state. In order to evaluate 
the EOF for the two modes initially excited in a TWB state, we can obtain 
the expression of the covariance matrix at time $t$.  

\subsection{Analytic solution in the weak coupling limit}
We consider a two-mode Gaussian states, that is, those states characterized 
by a Gaussian characteristic function $\chi_0(\Lambda)=\exp\{-\frac{1}{2}\Lambda^{T}
\sigma_0\Lambda-i\Lambda^{T}\bar{\mathbf{X}}_{in}\}$, where  $\sigma_0$ 
the initial  covariance matrix
\begin{equation}\label{CovMatzero}
\sigma_0=\left(
          \begin{array}{cc}
            \mathbf{A_0} & \mathbf{C_0} \\
            \mathbf{C^T_0} & \mathbf{B_0} \\
          \end{array}
        \right),
\end{equation}
where $\mathbf{A_0}= a\,{\mathbbm 1}$, $\mathbf{B_0}=b\,{\mathbbm 1}$,
$\mathbf{C_0}={\rm Diag}(c_1,c_2)$, and $a=b=\cosh(2r)$ and $c_1=-c_2=\sinh(2r)$
with $r>0$, and $\mathbbm 1$ the $2\times 2$ identity matrix.

Similar to Ref.  \cite{Vasile2009}  the interaction between oscillators and baths 
is bilinear in position and momentum, thus the evolution maintains the Gaussian 
character. The evolved state is a two-mode Gaussian state with mean 
and covariance matrix are described by 

\begin{align}
\bar{\mathbf{X}}_t&=e^{-\Gamma(t)/2}(R\oplus R)
\bar{\mathbf{X}}_{in} \\
\label{CovMatT}
\sigma_t&=e^{-\Gamma(t)}(R\oplus R)\sigma_0(R\oplus
R)^T+2(\bar{W}_t\oplus\bar{W}_t),
\end{align}
Using the weak coupling approximations described in Eqs.~\eqref{CharSol}--\eqref{eq:Mt} 
and replacing the matrix $\bar{W}_t$, the covariance matrix at time $t$ is now given by

\begin{equation}\label{CovMatT2}
\bmsigma_{t} = \left(
\begin{array}{c | c}
\bmA_{t} & \bmC_{t} \\ \hline \bmC_{t}^{T} & \bmA_{t}
\end{array}
\right).
\end{equation}
The analytic expression of the matrices $\bmA_{t}$ and $\bmC_{t} $ is given in the 
in the Appendix for both low and high temperature-reservoirs.

\par
In order to study in more details the entanglement dynamics in a non-Markovian 
channel, we need to specify the properties of the bosonic reservoirs. As a first model 
of the environment for the case of zero temperature Lorenztian spectral density has takes 
the form

\begin{equation}\label{den1}
J(\omega)=\frac {1} {\omega^{2}+\lambda^{2}},
\end{equation}
The parameter $\lambda$ represents the width of the distribution, which is connected to
the reservoir correlation time. In this model, the reservoir correlation time is given 
by the inverse of the system-reservoir interactions time, that is, $\tau_{R}=1/\lambda$. 
In order to make a comparative study, we could deal with any given form of the spectral 
density. But as a particular example, we use the following form of spectral density to 
specify the environments at high temperature and the spectral density is of the form

\begin{equation}\label{den2}
J(\omega)=\frac {\omega} {\omega^{2}+\lambda^{2}}.
\end{equation}
\par
The physical meaning when viewing the Lorentzian spectral density (\ref{den2}) as a mathematical 
vehicle to conveniently model a non-zero temperature bath correlation function with 
microscopically defined spectral density. We also concentrate on a particular class of 
entangled initial states, i.e. the TWB vacuum states, obtained by applying the two mode 
squeezing operator $S(\zeta)=\exp(\zeta a^{\dagger}b^{\dagger}-\zeta^{*}a b)$ 
to the vacuum state of the two modes, with $\zeta=r e^{i\varphi}$.  The TWB state is thus determined only by the squeezing parameter $r$ which also determines the initial amount of entanglement. 
An analytic expressions for the relevant time dependent coefficients given in Eq. \eqref{Express1} can be derived by 
approximating  in the high-$T$ and zero-$T$ limits, that is, for 
$2N(\omega)+1\approx\frac{2k_BT}{\hbar\omega}$ and $2N(\omega)+1\approx 1$, 
respectively (The mathematical steps are outlined in
Appendix A).

%%%%%%%%%
\subsection{Entanglement of Formation}

Entangled two-mode Gaussian states are a key resource for quantum 
technologies such as quantum cryptography, quantum computation 
and teleportation, so quantification of Gaussian entanglement is an important problem. 
A convenient and proper way of measuring quantum correlations in CV systems is by 
means of EOF  \cite{EoF,G:03}. EOF quantifies the minimum amount of two-mode 
squeezing needed to prepare an entangled state from a classical one. However, an 
analytical expression for EOF exists only for special cases, e.g., for two qubits \cite{Wootters} 
and for an arbitrary bimodal Gaussian state \cite{EoF2008} and finding a closed formula for 
an arbitrary state remains an open problem to this day. For simplicity, we will assume that 
the initial state is a symmetric bipartite Gaussian state with covariance matrix given by 
Eq. \eqref{CovMatzero}. As we mentioned above, when this state interacts with two 
identical  independent  reservoirs, the action of a Gaussian operation (any operation that 
transforms a Gaussian state into another Gaussian state) and the evolved covariance 
matrix is given by Eq. \eqref{CovMatT2}. The EOF for all symmetric 
Gaussian states corresponding to two modes is given by  \cite{G:03}
\begin{equation}\label{EoF}
E_{F} = (x_m + \mbox{$\frac12$}) \ln(x_m +\mbox{$\frac12$}) - (x_m -
\mbox{$\frac12$}) \ln(x_m - \mbox{$\frac12$}),
\end{equation}
with $x_m = (\tilde\kappa_{-}^{2} + 1/4)/(2 \tilde\kappa_{-})$,
$\tilde\kappa_{-}=\sqrt{(a_n - c_{+})(a_n - c_{-})}$ being the minimum
symplectic eigenvalue of the CM $\bmsigma_{t}$, and
\begin{align}
a_n &= \sqrt{I_1},\\
c_{\pm} &= \sqrt{\frac{I_1^2+I_3^2-I_4 \pm \sqrt{(I_1^2+I_3^2-I_4)^2
- (2 I_1I_3)^2}} {2 I_1}}, \label{c:pm}
\end{align}
where $I_1 = \det[\bmA_t]$, $I_3 = \det[\bmC_t]$ and $I_4 =
\det[\bmsigma_t]$ are the symplectic invariants of $\bmsigma_t$.

%%%%%%%%%
\section{The entanglement dynamics}\label{sec:ED}

Entanglement is not only a property of quantum mechanics but also 
of a crucial resource that allows certain quantum protocols to be more efficient 
than their classical counterpart \cite{ NieChu,Horo09}. However, the 
inevitable interaction between the quantum system and its environments leads to 
loss of entanglement. Studying this sort of model led to the discovery of the 
entanglement sudden death (ESD) phenomenon in which the entanglement 
between two qubits decays to zero in a finite time rather than asymptotically, has been 
predicted theoretically \cite{YuLettScien} and subsequently been verified experimentally 
\cite{Almeida}, indicating specific behavior of entanglement that differs from that of coherence. 
Reference \cite{Bellomo07} found that the entanglement can revive after 
some time interval of ESD and thus extends significantly the entangled time of the qubits. 
This remarkable phenomenon, which has been experimentally observed \cite{Xu10}, is 
physically due to the dynamical back action (that is, the non-Markovian effect) of the memory 
environments \cite{Bellomo07, Maniscalco08}. However, in many cases the finite extension 
of the entangled time is not enough and thus it is desired to preserve a significant fraction 
of the entanglement in the longtime limit. Indeed, it has shown \cite{Bellomo} that some 
noticeable fraction of entanglement can be obtained by engineering structured environment 
such as a photonic bandgap materials \cite{John90,John94,Berihu21}.  
\par 
In order to investigate the decoherence effect in the entanglement dynamics induced 
by the environment, a specification of the spectral density $J(\omega)$ of the environment 
is required. We have briefly investigated the parameters influencing the time evolution 
are the reservoir temperature, the form of the reservoir spectrum and the supporting 
mode $\omega_0$, resulting from the combination of the reservoir spectrum. We will 
 study the effect of these parameters separately in the following dealing with the dynamics 
 for  $T=0$, for high-$T$ reservoirs, and a comparison with different reservoir spectra. 
 We are also interested in finding the influence of how the initial state of the system, in 
 particular, on the initial squeezing parameter $r$ of TWBs on the validity of the secular 
 approximation.
 \par
 An interesting problem in this context, is entanglement dynamics for various types of 
 structured reservoirs and for different reservoir temperatures.
%%%%%%%%%%%%%
%\subsection{Zero-temperature regime} slowing down the loss of entanglement.
It is well known from previous studies on open quantum systems that interacting 
with zero-temperature reservoirs we expect on the one hand slowing down the loss 
of entanglement \cite{PraBec2004} and on the other hand more pronounced the non-Markovian 
features \cite{Alicki}, with respect to the $T\neq 0$ case. 
%We will have a closer look at these general fe atures of the dynamics in Sec. \ref{sec:ED} and focus 
%here on the validity of the secular approximation.
\par 

\begin{figure}[h!]
\includegraphics[width=0.49\columnwidth]{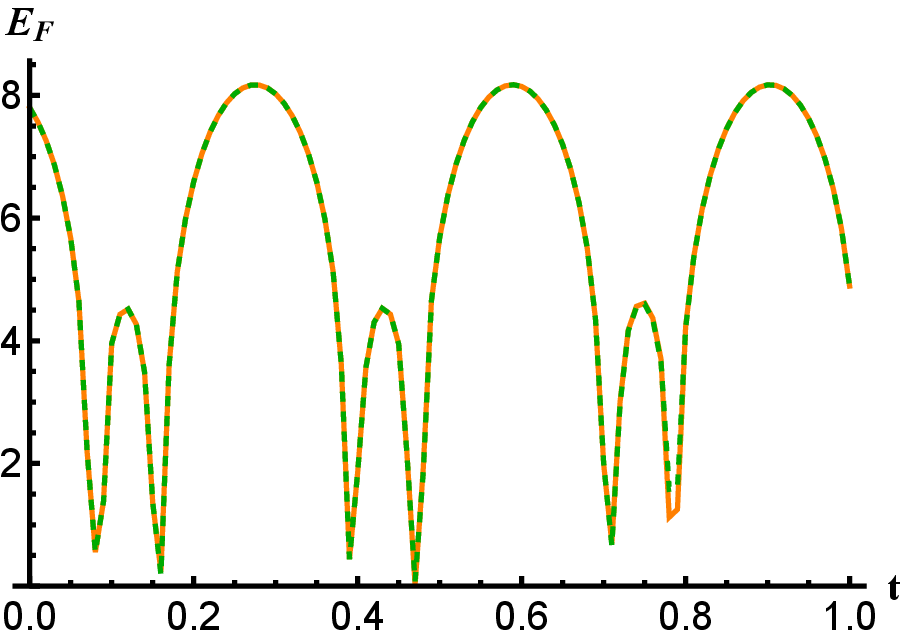}
\includegraphics[width=0.49\columnwidth]{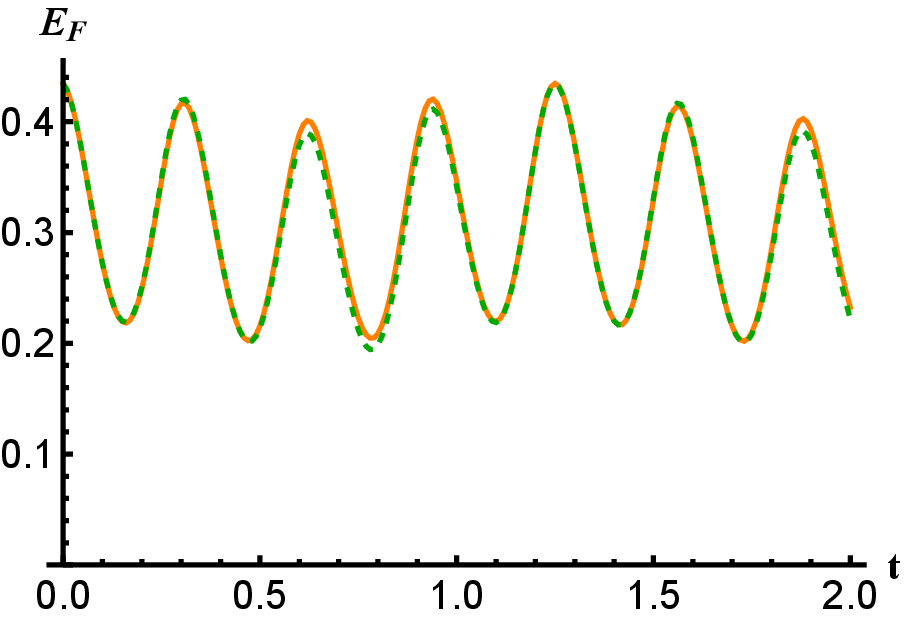}

\caption{(Colors online)  Dynamics of exact EoF (solid red line) and the secular approximate 
(dashed blue line) as a function of $t$, for the Lorentzian spectral density model (\ref{den1}) at 
zero temperature for $\alpha=0.1$, $\lambda=0.1$, with $\omega_0=5$, (a) $r=2$ and (b) $r=0.2$. 
The two curves almost overlap perfectly.}\label{fig:1}
\end{figure}

To address this problem, we now proceed with a reservoir of the form (\ref{den1}) 
at zero temperature Lorentzian bath with $\omega_{0}=5$ and look at the dynamics of a TWB 
with a different amount of initial entanglement, $r$. In Fig. \ref{fig:1} (left panel),
we show an example of entanglement evolution characterized by oscillations, sudden death 
and revivals. Fig. \ref{fig:1} (right panel), shows that the evolution of the entanglement 
persist longer for a longer time at $r=0.2$ and appear oscillations even as the state 
approaches the ground state. As one would expect, the exact and the secular 
approximated dynamics completely agree in this situation. We have also carefully 
examined the dynamical behavior for other chosen parameters of $\omega_{0}$ and 
of the initial squeezing parameter $r$, we observe that there is quite general property 
of the system. Therefore, for a general description of bimodal CV quantum systems 
interacting with zero temperature reservoirs, the secular approximation 
can always be performed well and we always neglect the effects of the nonsecular terms. 
We realize from this that the secular terms \eqref{SecCoeff} are temperature dependent through 
the diffusion coefficients $\Delta(t)$ and $\Pi(t)$, and at $T=0$ their contribution is 
rather small. 
\par
\begin{figure}[h!]
\includegraphics[width=0.49\columnwidth]{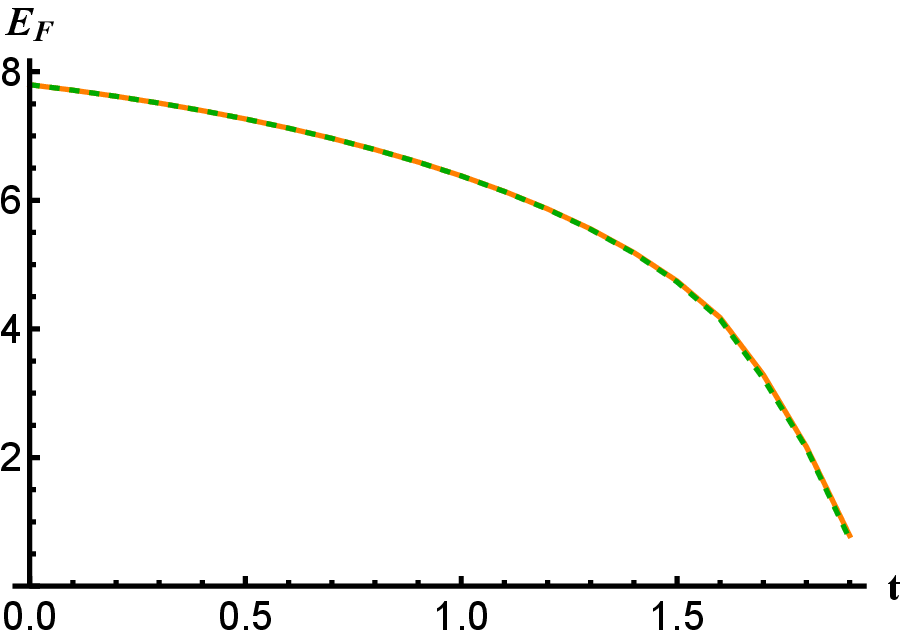}
\includegraphics[width=0.49\columnwidth]{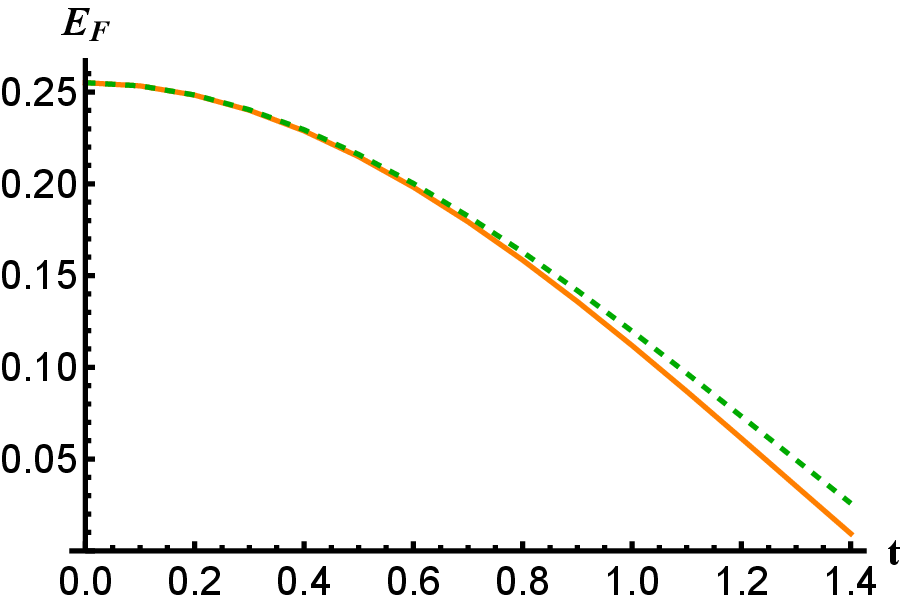}
\caption{(Colors online)  The exact EoF (solid orange line) and the secular approximate dynamics 
(dashed green line) of the $E_F$ as a function of $t$, for the Lorentzian spectral density model (\ref{den1}) at 
zero temperature for $\alpha=0.1$, $\lambda=0.1$, with $\omega_0=0.2$, 
(a) $r=2$ and (b) $r=0.1$.  The two curves almost overlap perfectly for higher values of squeezing.}\label{fig:11}
\end{figure}

For intermediate values of the parameter $\omega_0$, $\omega_{0} \le 1$, we observe 
a stronger dependence on the initial value of the entanglement. In Fig.  \ref{fig:11}, we plot 
the time evolution of the EOF for the entanglement dynamics calculated using the secular 
approximated solution and the exact solution for two different initial TWB states.  For a range 
of higher values of the initial entanglement, the entanglement dynamics are robust to 
decoherence.

%%%%%%%%%%%
%\subsection{High-temperatures regime}

In the same way as above, we consider the high-temperature limit $k_BT\gg\hbar\omega_0$, that is, 
when the classical thermal energy $k_BT$ is much larger than the typical energies exchanged in 
our system. In the following analysis we choose a temperature such that $k_BT/\hbar=100$, thus 
we can examine scenarios in which $ \omega_0\geq0.1$.
\par
\begin{figure}[h!]
\includegraphics[width=0.49\columnwidth]{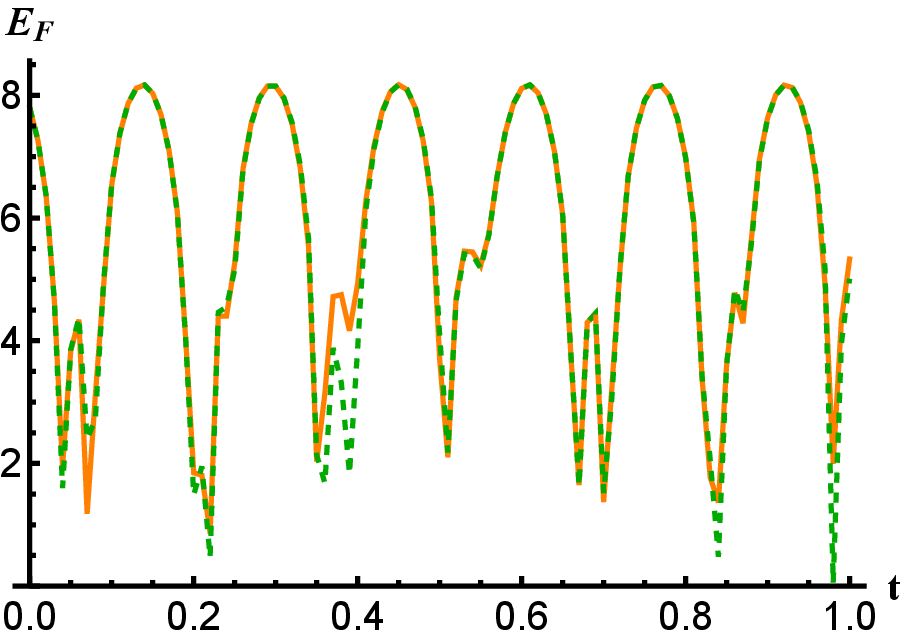}
\includegraphics[width=0.49\columnwidth]{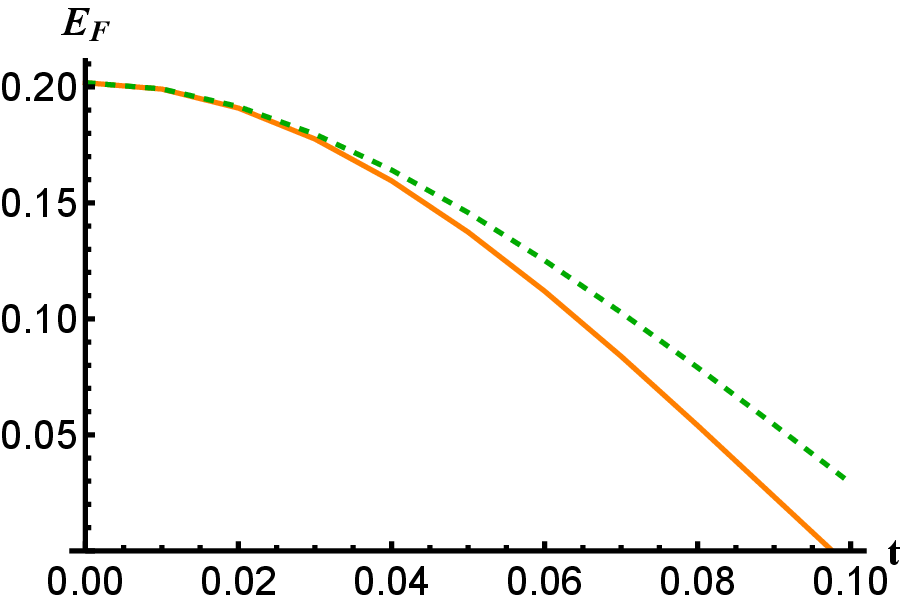}
\caption{(Colors online) Comparison between the exact 
EoF dynamics (solid orange line) and the secular approximate dynamics (dashed green line) 
as a function of $t$, with $\omega_0=10$, (a) $r=2$ and (b) $r=0.01$. We set 
$k_BT/\hbar=100$, $\lambda=0.1$, $\alpha=0.1$.}
\label{fig:2}
\end{figure}

It will turn out that the behavior of the time evolution of entanglement in CV quantum channels 
in the class of Ohmic-like spectral distribution with an exponential cutoff function has been 
studied in detail in \cite{Vasile2009}. We can however summarize the most 
important features for the non-Markovian short time dynamics of entanglement in CV quantum 
channels in the case of the spectral distribution of the form (\ref{den2}) and $\omega_0=10$. 
In Fig. \ref{fig:2}, we plot the important features of the time evolution of the EOF calculated 
using both the secular approximated solution and the exact solution in the regime 
$\omega_0\geq1$ for two different initial TWB states. Figure  \ref{fig:2} (left panel) shows that for 
higher values of the initial entanglement, that is, for larger values of $r$, the entanglement 
dynamics exhibits the behavior of oscillations. For small initial entanglement ($r=0.01$), 
where the entanglement dynamics is significantly changed as we see from Fig.  \ref{fig:2} (right panel). 
One may also see that the exact and the secular approximated dynamics sensibly agree 
for $r=2$ (higher initial entanglement) during the initial time but fails after $t\ge0.1$ and 
for small value of the initial entanglement $r=0.01$, the secular approximation does not work well. 
Furthermore, both the secular approximation and the exact results show oscillations 
for higher values of the initial entanglement as clearly visible in the dynamics. 
From previous study \cite{Vasile2009}, one would have conclude that the entanglement 
dynamics in the high temperature case is affected by the secular approximation. 
Our results, however, clearly show on the contrary that the entanglement dynamics 
at high temperature and for smaller values of initial entanglement, the secular terms 
predicts much longer disentanglement time.

\par
\begin{figure}[h!]
\includegraphics[width=0.49\columnwidth]{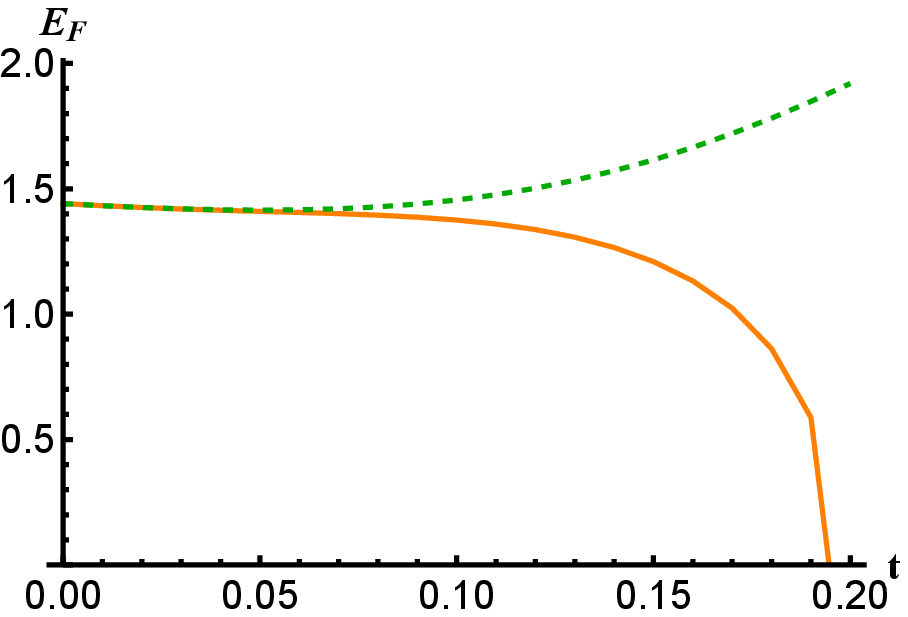}
\includegraphics[width=0.49\columnwidth]{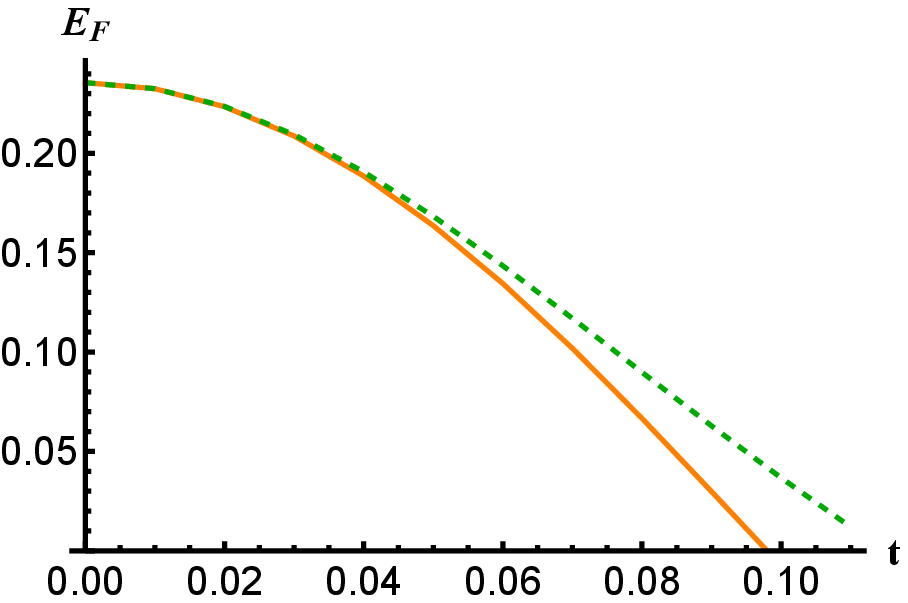}
 \caption{(Colors online) Dynamics of  the exact EoF 
dynamics (solid orange line) and the secular approximate dynamics (dashed green line) 
as a function of $t$, with $\omega_0=0.15$, (a) $r=1$ and (b) $r=0.08$. We 
set $k_BT/\hbar=100$, $\lambda=0.1$, $\alpha=0.1$.}
\label{fig:3}
\end{figure}

Having analyzed the differences in the entanglement dynamics for values of the parameters 
$\omega_0\geq1$, we now focus on the intermediate values of the parameters $\omega_0$,  
$\omega_0\leq1$, we observe a strong dependence on the initial value entanglement. In 
Fig.  \ref{fig:3}, we study the dependence of the initial value entanglement and shows that for 
both initial conditions the entanglement oscillations are strongly suppressed. As one would expect, 
 the secular approximations for the high-$T$ dynamics affected by this approximation. On the other 
 hand, for both initial values of the entanglement the amplitude of the secular approximations 
 increases. 

 \par
 So far, we have studied the entanglement dynamics, as measured by the entanglement of formation, 
 for the two modes initially excited in a TWB for a single Lorenztian spectral distribution of 
 the form given in (\ref{den1}) and (\ref{den2}). The phenomenon of entanglement sudden death has clearly provoked much theoretical 
interest, and it is related to another question that is both interesting from a theoretical 
perspective and clearly of great practical importance: how can one protect a system from 
disentanglement? Here we do not propose any active scheme for protecting entanglement 
but rather consider what initial amount of entanglement and the ranges of $\omega_0$ tend 
to minimize the loss of entanglement or safeguard entanglement once it has been dynamically 
generated. Of particular interested is avoiding ESD.
Another intriguing aspect is now how the entanglement preservation is influenced by the 
effect of temperature on the non-Markovian dynamics. As shown in Figure (\ref{fig:3}), we sharply 
note that the exact solution containing the nonsecular terms exhibits ESD. On one hand, the 
disentanglement time predicted by the exact and secular results increases with increasing 
values of initial entanglement, that is, for larger values of $r$. Meanwhile, the entanglement 
oscillations is sustained for the longest time for higher values of the initial entanglement. Finally, 
as shown in Figure (\ref{fig:1}), we investigate the effects of higher value of $\omega_0$ for 
different values of initial entanglement. It is interesting to note that the time evolution of the entanglement 
dynamics has significant differences for different values of initial entanglement $r$. For a range of small 
values of $r$, one can clearly see that the the entanglement exhibits the behavior of oscillation and is sustained
for longer time. Also in this case one can clearly see that the exact and secular approximated dynamics sensibly agree.
We should mention that the entanglement oscillations have also have also been found for two uncoupled 
harmonic oscillators interacting with two independent reservoirs \cite{Maniscalco}. In the following, we investigate 
the difference in the loss of entanglement due to the choice of the amount of entanglement in the initial TWB state, given by the value of squeezing parameter $r$ and due to different reservoir spectra. We have seen that the non-Markovian revivals of entanglement occur for intermediate value of the squeezing parameter $r$ (right panel in the high-T limit) and with the state being AL (left panel for the zero-T case). In all cases there are entanglement oscillations and ESD, while an entanglement revival is also present for higher values of $\omega_0$ and the initial amount of entanglement $r$. 
\begin{figure}[h!]
\includegraphics[width=0.49\columnwidth]{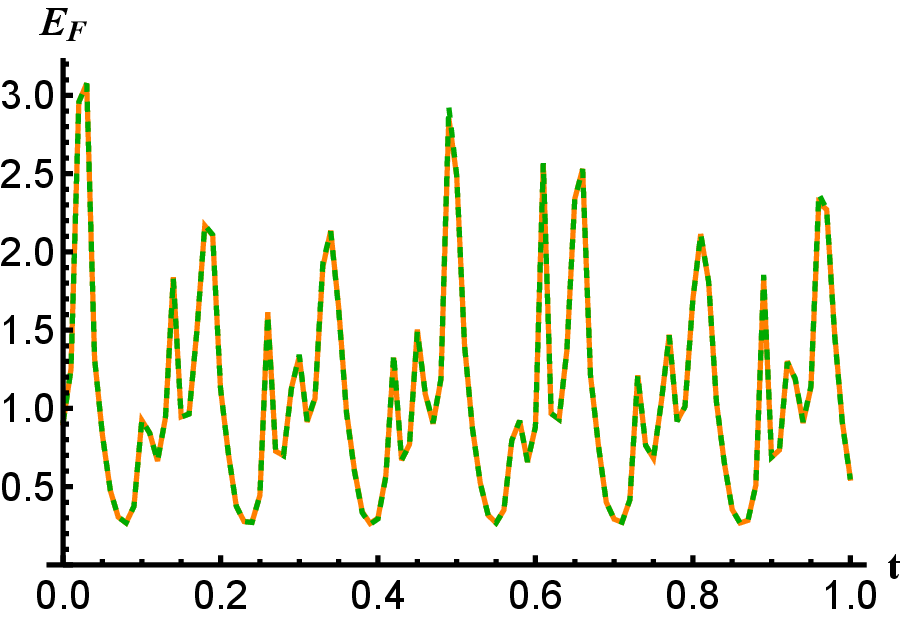}
\includegraphics[width=0.49\columnwidth]{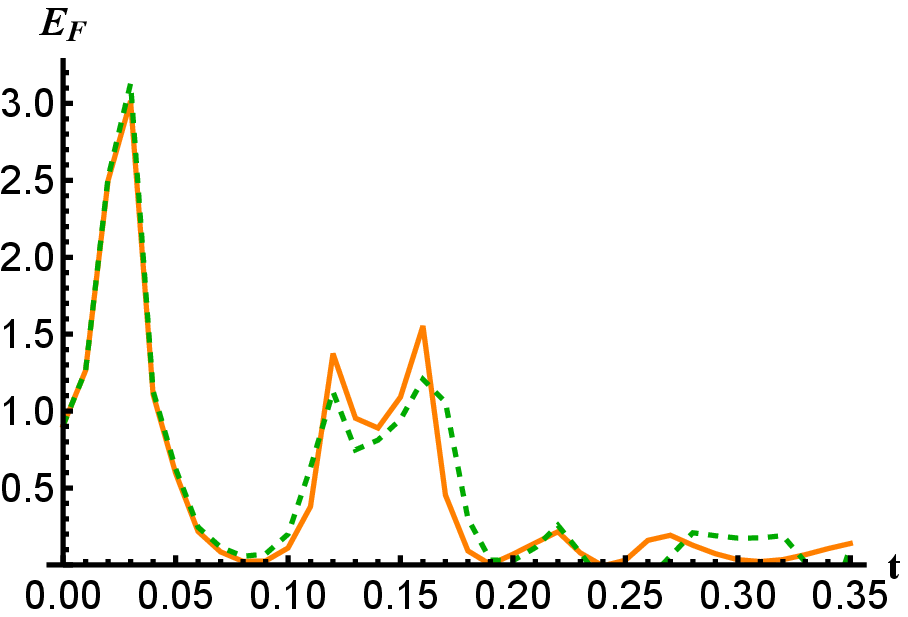}
 \caption{(Colors online) Dynamics of  the exact EoF 
dynamics (solid orange line) and the secular approximate dynamics (dashed green line) 
as a function of $t$, with $\omega_0=10$, $r=0.6$ for the Lorentzian environment reservoir 
of the form (\ref{den1}) (Left) and for the high-$T$ limit (\ref{den2}). We 
set $k_BT/\hbar=100$, $\lambda=0.1$, $\alpha=0.1$.}
\label{fig:4}
\end{figure}

%%%%%%%%%%%%%%
\section{Summary and outlook}\label{s:concl}
In the present work, we have studied the entanglement dynamics in a model consisting of 
two different Lorentzian spectral densities in non-Markovian environments as measured 
by the entanglement of formation, for the two modes initially excited in a twin-beam state 
in terms of phenomena such as dynamical entanglement generation, entanglement 
sudden death, and an ``always alive'' (AL) phenomenon where entanglement once 
present never goes to zero at any time. A similar work has also been 
studied the evolution of entanglement in bimodal continuous variable quantum systems 
with two independent structure reservoir in thermal equilibrium \cite{Maniscalco,Vasile2009,Vasile2010}.   
\par
We investigated the entanglement dynamics as a function of the reservoir spectrum, the temperature 
and the initial amount of entanglement. In our first Lorentzian bath (\ref{den1}), we considered higher 
values of $\omega_0$ in the zero-temperature case showed that the evolution of the entanglement persists longer 
for a longer time and appear oscillations for intermediate values of the initial entanglement. Moreover, 
for $\omega_0<1$, the entanglement dynamics are completely robust against decoherence. Note also 
that the secular and exact dynamics almost coincide for higher initial entanglement. In the second 
Lorentzian model (\ref{den2}) at high-temperatures we see that the non-Markovian entanglement 
oscillations persists for longer time for higher values of the initial entanglement. 
We also see from Fig. \ref{fig:2} (right panel) that the exact solution in the 
high-$T$ dynamics has a shorter disentanglement time. At high-temperature and intermediate 
values of the parameters $\omega_0<1$, ESD appears for each value of the initial entanglement 
for the exact dynamics and entanglement predicts longer for the secular solutions. We also find that 
entanglement is more robust in the case of zero-temperature Lorentzian bath while it vanishes 
faster for the high-temperature Lorentzian bath. Also Gaussian entanglement oscillations \cite{Maniscalco} 
are evident for the higher values of the reservoir spectrum and intermediate value of the squeezing parameter, 
a result independent, qualitatively, on the temperature of the bath. 

Recently, there have been a lot of interest in the entanglement dynamics in CV quantum channels, both for common
and independent reservoirs. The result presented in this article demonstrate that careful manipulation 
of the non-Markovian short time dynamics of entanglement in different physical scenarios. This means that, 
in certain situations, the effects of different reservoirs on the time evolution of entanglement in CV quantum channels, 
with the potential to pave the way to real-scale quantum-enhanced devices \cite{Bylicka}.

\section*{Acknowledgments}
\noindent I would like to acknowledge Matteo Paris and Stefano Olivares for helpful discussions. 
This work has been supported by Khalifa University through project no. 8474000358 (FSU-2021-018).
%%%%%%%%%%%%%%%%%
%\begin{widetext}
\appendix

\section{Time dependent coefficients at the second order in $\alpha$}

The time-dependent coefficients of the master equation given in \eqref{QBMme}, 
n the case of thermal reservoirs in the weak coupling regime  ($\alpha\ll 1$) are given by

\begin{subequations}\label{Express1}
\begin{align}
\Delta(t)&=\alpha^2\!\!\int_0^t\!\! ds \! \int_0^{+\infty}\!\!\!\!\!\!\!\!
\!d\omega
J(\omega)[2N(\omega)+1]\cos(\omega s)\cos(\omega_0 s), \\
\Pi(t)&=\alpha^2\!\!\int_0^t\!\! ds \! \int_0^{+\infty}\!\!\!\!\!\!\!\!
\!d\omega
J(\omega)[2N(\omega)+1]\cos(\omega s)\sin(\omega_0 s), \\
\gamma(t)&=\alpha^2\!\!\int_0^t\!\! ds \! \int_0^{+\infty}\!\!\!\!\!\!\!\!
\!d\omega
J(\omega)\sin(\omega s)\sin(\omega_0 s), \\
r(t)&=\alpha^2\!\!\int_0^t\!\! ds \!
\int_0^{+\infty}\!\!\!\!\!\!\!\! \!d\omega J(\omega)\sin(\omega
s)\cos(\omega_0 s),
\end{align}
\end{subequations}
where $N(\omega)=[\exp(\hbar\omega/k_BT)-1]^{-1}$ is the mean
number of photons with frequency $\omega$, and $\alpha$ is the dimensionless
system-reservoir coupling constant, while $J(\omega)$ defines the spectral distribution 
of the environment.\\

%%%%%%%%%%%%%%
\noindent For the Lorentzian spectral density of the form $J(\omega)=\frac {1} {\omega^{2}+\lambda^{2}}$, 
in the low $T$ limit ($2N(\omega)+1\approx 1$) and we made use of the following special 
mathematical functions \cite{Abramowitz}:
\begin{equation}
\text{Ei}(z)=-\int_{-z}^{+\infty}\frac{e^{-t}}{t} dt \quad \text{Si}(z)=\int_{0}^{z}\frac{\sin t}{t} dt. \nonumber
\end{equation}

\begin{align}
\gamma(t)&=\frac{\alpha^2}{\omega_{0}^{2}+\lambda^{2}}\biggl\{\biggl(\mathrm{Si}(\omega_{0}t)\\  \nonumber
&+\frac{\pi}{2}\biggl)+\frac{1}{2}\frac{\omega_{0}}{\lambda}\cos(\omega_{0}t)\biggl(e^{\lambda t}\mathrm{Ei}(-\lambda t)-e^{-\lambda t}\mathrm{Ei}(\lambda t)\biggl)\\ \nonumber
&-\frac{1}{2}\sin)\omega_{0}t)\biggl(e^{\lambda t}\mathrm{Ei}(-\lambda t)+^{-\lambda t}\mathrm{Ei}(\lambda t)\biggl)\biggl\}
\end{align}

\begin{equation}\begin{split}
\Delta_0(t)=\frac{\pi\alpha^2}{2(\omega_{0}^{2}+\lambda^{2})}\biggl\{e^{-\lambda t}\biggl(\frac{\omega_{0}}{\lambda}\sin(\omega_{0} t)-\cos(\omega_{0} t)\biggl)+1\biggl\}
\end{split}\end{equation}
\begin{align}
\Pi_0(t)&=\frac{\pi\alpha^2}{2(\omega_{0}^{2}+\lambda^{2})}\biggl\{\frac{\omega_{0}}{\lambda}\biggl(1-\cos (\omega_{0}t)e^{-\lambda t}\biggl)\\ \nonumber
&-\sin(\omega_{0}t)e^{-\lambda t}\biggl\}
\end{align}

\noindent For the Lorentzian spectral density of the form $J(\omega)=\frac {\omega} {\omega^{2}+\lambda^{2}}$ in the high-$T$ limit 
($2N(\omega)+1\approx\frac{2k_BT}{\hbar\omega}$), we have

\begin{align}
\gamma(t)&=\frac{\alpha^2\beta}{\omega_{0}^{2}+\lambda^{2}}\biggl\{\biggl(\mathrm{Si}(\omega_{0}t)+\frac{\pi}{2}\biggl)\\  \nonumber
&+\frac{1}{2}\frac{\omega_{0}}{\lambda}\cos(\omega_{0}t)\biggl(e^{\lambda t}\mathrm{Ei}(-\lambda t)-e^{-\lambda t}\mathrm{Ei}(\lambda t)\biggl)\\ \nonumber
&-\frac{1}{2}\sin)\omega_{0}t)\biggl(e^{\lambda t}\mathrm{Ei}(-\lambda t)+^{-\lambda t}\mathrm{Ei}(\lambda t)\biggl)\biggl\}
\end{align}

\begin{equation}\begin{split}
\Delta_T(t)=\frac{\pi\alpha^2\beta}{2(\omega_{0}^{2}+\lambda^{2})}\biggl\{e^{-\lambda t}\biggl(\frac{\omega_{0}}{\lambda}\sin(\omega_{0} t)-\cos(\omega_{0} t)\biggl)+1\biggl\}
\end{split}\end{equation}

\begin{align}
\Pi_T(t)&=\frac{\pi\alpha^2\beta}{2(\omega_{0}^{2}+\lambda^{2})}\biggl\{\frac{\omega_{0}}{\lambda}\biggl(1-\cos (\omega_{0}t)e^{-\lambda t}\biggl)\\ \nonumber
&-\sin(\omega_{0}t)e^{-\lambda t}\biggl\}
\end{align}

\noindent We do not provide the analytic expression of $r(t)$ because its contribution is negligible to the solution in the weak coupling limit.
\section{The master equation solution}
The time evolution of the characteristic function in the decoherence model (\ref{CharSol}) with expressions (\ref{mat1}), (\ref{mat2}), and 
(\ref{eq:Mt}), the following functions appear after explicit calculations
  \begin{subequations}
\label{SecCoeff}
\begin{align}
&\Gamma(t)=2\int_0^t \gamma(s)ds\\
&\Delta_{\Gamma}(t)=e^{-\Gamma(t)}\int_0^t e^{\Gamma(s)}\Delta(s)ds\\
&\Delta_{co}(t)=e^{-\Gamma(t)}\int_0^t
e^{\Gamma(s)}\Delta(s)\cos[2\omega_0(t-s)]ds,\\
&\Delta_{si}(t)=e^{-\Gamma(t)}\int_0^t e^{\Gamma(s)}\Delta(s)\sin[2\omega_0(t-s)]ds,\\
&\Pi_{co}(t)=e^{-\Gamma(t)}\int_0^t
e^{\Gamma(s)}\Pi(s)\cos[2\omega_0(t-s)]ds,\\
&\Pi_{si}(t)=e^{-\Gamma(t)}\int_0^t
e^{\Gamma(s)}\Pi(s)\sin[2\omega_0 (t-s)]ds.
\end{align}\end{subequations}

The explicit expression of the coefficients above depends on both the reservoir spectral 
density and the temperature. In order to find the analytic expression of the covariance matrix (\ref{CovMatT2}), 
we need to apply definition (\ref{CharSol}) and the properties of the Gaussian characteristic function to Eq. (\ref{CovMatT}). 
To do so, we find the analytical expressions of the covariance matrices at time $t$ can be written as   

\begin{align}
\bmA_{t} ={} \bmA_{0} e^{-\Gamma}
+ \left(
\begin{array}{cc}
\Delta_{\Gamma} + (\Delta_{\rm co} - \Pi_{\rm si}) &-(\Delta_{\rm si} - \Pi_{\rm co}) \\
-(\Delta_{\rm si} - \Pi_{\rm co}) & \Delta_{\Gamma} - (\Delta_{\rm co} - \Pi_{\rm si})
\end{array}
\right),
\end{align}\\
\begin{align}
\bmC_{t} = \left(
\begin{array}{cc}
c\, e^{-\Gamma}\, \cos(2\omega_0 t) &
c\, e^{-\Gamma}\, \sin(2\omega_0 t) \\
c\, e^{-\Gamma}\, \sin(2\omega_0 t) & -c\, e^{-\Gamma}\,
\cos(2\omega_0 t)
\end{array}
\right).
\end{align}

%\end{appendix}

%%%%%%%%%%%%%%
\par
%\section*{References}


\begin{thebibliography}{9}


\bibitem{NieChu} M. A. Nielsen, I.L. Chuang, \emph{Quantum Computation and Quantum
Information}, (Cambridge University Press, Cambridge, 2000)
\bibitem{Horo09} R. Horodecki, P. Horodecki, M. Horodecki, and K. Horodecki, Rev. Mod. Phys. {\bf81}, 865 (2009).


\bibitem{Walls} D. F. Walls and G. J. Milburn, {\em Quantum Optics} (Springer, Berlin, Heidelberg, 1994).
\bibitem{Breuer} H.-P. Breuer and F. Petruccione, {\em The Theory of Open Quantum Systems}
(Oxford University Press, Oxford, 2002).
\bibitem{Weiss} U. Weiss, {\em Quantum Dissipative Systems}, (World Scientific, Singapore,2008).
\bibitem{Lind} G. Lindblad, Commun. Math. Phys, {\bf 48}, 119 (1976).
\bibitem{Zurek} W. H. Zurek, Rev. Mod. Phys. {\bf 75}, 715 (2003). 
\bibitem{Lu} B. L. Hu, J. P. Paz, and Y. Zhang, Phys. Rev. D {\bf 45}, 2843 (1992).

\bibitem{Imamoglu} A. Imamoglu, Phys. Rev. A {\bf 50}, 3650 (1994).
\bibitem{Garrya} B. M. Garryway, Phys. Rev. A {\bf 55}, 2290 (1997).
\bibitem{Garry} B. M. Garryway, Phys. Rev. A {\bf 55}, 4636 (1997).
\bibitem{Mazzola09} L. Mazzola, S. Maniscalco, J. Piilo, K.-A. Suominen, and B. M. Garryway, Phys. Rev. A {\bf 79}, 042302 (2009).
\bibitem{Mazzola} L. Mazzola, S. Maniscalco, J. Piilo, K.-A. Suominen, and B. M. Garryway,  Phys. Rev. A {\bf 80}, 012104 (2009).


\bibitem{Richard} R. Hartmann and W. T. Strunz, Phys. Rev. A {\bf101}, 012103 (2020).
\bibitem{Holevo} A. S. Holevo and R. Werner, Phys. Rev. A {\bf 63}, 032312 (2001).
\bibitem{Serafini} A. Serafini {\em et al}., J.  Opt. B Quantum Semiclassical Opt. {\bf 7}, R19 (2005).
\bibitem{Ban} M. Ban, J. Phys. A {\bf 39}, 1927 (2006); Phys. Lett. A {\bf 359}, 402 (2006).
\bibitem{McAneney} H. McAneney {\em et al}., J. Mod. Opt. {\bf 52}, 935, 0950  (2005) .
\bibitem{Maniscalco} S. Mansicalco, S. Olivares, and M. G. A. Paris, Phys. Rev. A {\bf 75}, 062119 (2007).
\bibitem{Jhn} J. H. An, M. Feng, and W. M. Zhang, arXiv:0705.2472.
\bibitem{Kl} K. -L. Liu and H.-S. Goan, Phys. Rev. A {\bf 76}, 022312 (2007).
\bibitem{Jhn1} J. H. An and W. M. Zhang, Phys. Rev. A {\bf 76}, 042127 (2007).
\bibitem{Jhn2} J-H. An, Y. Yeo, W-M Zhang, and C. H. Oh, arXiv:0811.1309.


\bibitem{Matteo} M. G.  A. Paris, J. Opt. B, {\bf 4}, 442 (2000).
\bibitem{Serafini1}  A. Serafini, F. Illuminati, M. G. A. Paris and S. De Siena, Phys. Rev. A {\bf69}, 022318 (2004).   


\bibitem{Braunstein}
S. L. Braunstein and P. van Loock, Rev. Mod. Phys. {\bf 77}, 513 (2005).

\bibitem {Vasile2009}  R. Vasile, S. Olivares, M. G. A. Paris and S. Maniscalco, Phys. Rev. A {\bf 67}, 062324 (2009). 

\bibitem {Vasile2010}  R. Vasile, P. Giorda, S. Olivares, M. G. A. Paris and S. Maniscalco, Phys. Rev. A {\bf82}, 012312 (2010).  

\bibitem {Vasile2011} R. Vasile, S. Maniscalco, M. G. A. Paris, H.-P. Breuer and J. Piilo, Phys. Rev. A {\bf84}, 052118 (2011). 
\bibitem {Maniscalco13} P. Haikka, T. H. Johnson, and S. Maniscalco, Phys. Rev. A {\bf87},010103(R) (2013). 

\bibitem {Bina2018} M. Bina, F. Grasselli, and M. G. A. Paris, Phys. Rev. A {\bf97}, 012125 (2018).  


\bibitem {Alex} A. W. Chin, S. F. Huelga, M. B. Plenio, Phys. Rev. Lett. {\bf109}, 233601 (2012)

\bibitem {Claudia} C. Benedetti, F. S.  Sehdaran, M. H. Zandi, and M. G. A. Paris, 
Phys. Rev. A {\bf97}, 012126 (2018).

\bibitem {Albarelli} F. Albarelli, M. A. C. Rossi, D. Tamascelli, and M. G. Genoni, Quantum {\bf2}, 110 (2018).

\bibitem {Fahimeh} F. S. Sehdaran, M. Bina, C. Benedetti, M. G. A. Paris, Entropy{\bf 21}, 486 (2019).

\bibitem {Dario} D. Tamascelli, C. Benedetti, H.-P. Breuer, and M. G. A. Paris, New J. Phys. {\bf22}, 083027 (2020).

\bibitem {Gebbia} F. Gebbia, C. Benedetti, F. Benatti, R. Floreanini, M. Bina, and M. G. A. Paris, 
Phys. Rev. A {\bf101}, 032112 (2020).

\bibitem {Candeloro} A. Candeloro and M. G. A. Paris, Phys. Rev. A {103}, 012217 (2021).

\bibitem {Ze} Z-Z Zhang and W. W, Phys. Rev. Res.  {\bf3}, 043039 (2021).

\bibitem {Vasile2011a} R. Vasile, S. Olivares, M. G. A. Paris and S. Maniscalco, Phys. Rev. A {\bf83}, 042321 (2011). 

\bibitem {Bylicka} B. Bylicka, D. Chru\'{s}ci\'{n}ski and S. Maniscalco, Scientific Reports, {\bf4}, 5720 (2014). 


\bibitem{EoF} C. H. Bennett, D. P. DiVincenzo, J. A. Smolin and W. K.
Wootters, Phys. Rev. A {\bf 54}, 3824 (1996).

\bibitem{G:03} G. Giedke, M. M. Wolf, O. KruÂger, R. F. Werner, and J. I. Cirac, Phys. Rev. Lett. {\bf 91}, 107901 (2003).

\bibitem{EoF2008} P. Marian and T. A. Marian, Phys. Rev. Lett. {\bf 101},
220403 (2008).

\bibitem{Liu} B.-H. Liu and L. Li and Y.-F. Huang and C.-F Li and G.-C. Guo and E.-M. Laine and
H.-P. Breuer and J. Piilo, Nat. Phys. {\bf7}, 931 (2011).

\bibitem{Tang} J.-S. Tang and C.-F. Li and Y.-L. Li and X.-B. Zou and G.-C. Guo and H.-P. Breuer
and E.-M. Laine and J. Piilo, Europhys. Lett. {\bf97}, 10002 (2012).

\bibitem{Intra2003} F. Intravaia, S. Maniscalco, and A.
Messina, Phys. Rev. A \textbf{67}, 042108 (2003).

\bibitem{Wootters}  W. K. Wootters,  Phys. Rev. Lett. {\bf 80}, 2245?2248 (1998).


\bibitem{YuLettScien}
T. Yu and J. H. Eberly, Phys. Rev. Lett. {\bf93}, 140404 (2004); Science {\bf 323 }, 598 (2009).
\bibitem{Almeida} M. P. Almeida, F. de Melo, M. Hor-Meyll, A. Salles, S. P.
Walborn, P. H. S. Ribeiro, and L. Davidovich, Science {\bf316}, 579 (2007).


\bibitem{Bellomo07} B. Bellomo, R. LoFranco, and G. Compagno, Phys. Rev. Lett. {\bf99}, 
160502 (2007); Phys. Rev. A {\bf77}, 032342 (2008).

\bibitem{Xu10} J.-S. Xu, C.-F. Li, M. Gong, X.-B. Zou, C.-H. Shi, G. Chen, and
G.-C. Guo, Phys. Rev. Lett. 104, 100502 (2010).

\bibitem{Maniscalco08} S. Maniscalco, F. Francica, R. L. Zaffino, N. L. Gullo, and F. Plastina, Phys. Rev. Lett. {\bf100}, 090503 (2008). 

\bibitem{Bellomo} B. Bellomo, R. L Franco, S. Maniscalco, and G. Compagno, Phys. Rev. A {\bf78}, 060302(R) (2008).

\bibitem{John90} S. John and J. Wang, Phys. Rev. Lett. {\bf64}, 2418 (1990).
\bibitem{John94} S. John and T. Quang, Phys. Rev. A {\b50}, 1764 (1994).

\bibitem{Berihu21} B. Teklu, {\em et al} (in preparation).

\bibitem{PraBec2004} J. S. Prauzner-Bechcicki, J.
Phys. A \textbf{37}, L173 (2004).


\bibitem{Alicki}
R. Alicki, M. Horodecki, P. Horodecki, and R. Horodecki, Phys. Rev. A {\bf 65}, 062101 (2002).

\bibitem{Abramowitz}  M. Abramowitz, {\em Handbook of Mathematical Functions}, edited by I. A. Stegun Dover 
Publication, New York, (1965.)



\end{thebibliography}
\end{document}